# Integrated Mobile Solutions in an Internet-of-Things Development Model

Issam Damaj and Safaa Kasbah

**Abstract** The Internet-of-Things (IoT) is a revolutionary technology that is rapidly changing the world. IoT systems strive to provide automated solutions for almost every life aspect; traditional devices are becoming connected, ubiquitous, pervasive, wireless, context-aware, smart, controlled through mobile solutions, to name but a few. IoT devices can now be found in our apartments, places of work, cars, buildings, and in almost every aspect of life. In this investigation, we propose an IoT system Development Model (IDM). The proposed IDM enables the development of IoT systems from concept to prototyping. The model comprises concept refinement pyramids, decision trees, realistic constraint lists, architecture and organization diagrams, communication interface patterns, use cases, and menus of analysis metrics and evaluation indicators. The investigation confirms that the proposed model enjoys several properties, such as, clarity, conciseness, thoroughness, productivity, etc. The model is deployed for a variety of systems that belong to heterogeneous areas of application; the model is proven to be effective in application and successful in integrating mobile solutions. This chapter includes the presentation of the IDM sub-models, the reasoning about their usefulness, and the technical developments of several systems. The chapter includes thorough discussions, analysis of the model usability and application, and in-depth evaluations.

---

Issam Damaj
Electrical and Computer Engineering Department, American University of Kuwait, Salmiya, Kuwait, e-mail: idamaj@auk.edu.kw

Safaa Kasbah
Computer Science and Mathematics Department, Lebanese American University, Beirut, Lebanon, e-mail: safaa.kasbah@lau.edu.lb

## 1 Introduction

The quest for automating places of living, work, and services has ever existed. The aim of all times has been to reduce the amount of labor, saving energy, saving materials, improving quality of life, increasing security, enhancing safety, and more. In the old times, ancient societies had float-valve regulators, water clocks, oil lamps, water tanks, water dispensers, etc. The mechanical revolution brought us the much-needed locks, switches, pumps, shading devices, etc. The electrical revolution introduced relays, switches, programmable controllers, and more. Recently, embedded systems such as microcontrollers and programmable devices enabled the automation of an almost infinite list of gadgets, equipment, vehicles, tools, etc. In the 1990s, carrying out a home automation project means the sole responsibility to fabricate all interfacing, port expansion, and bus cards. In addition, the development should include all the soft interfaces. Although the electronic designs were somewhat available, easy to reach and off-the-shelf purchases were almost non-existent.

At present, a wide range of plug-and-play computer interfacing kits are made available by a variety of providers, such as, the Arduino Project, Phidgets, Raspberry Pi, Handy Kit, National Instruments (NI) systems, to name a few. The richness of present-day interfacing gadgets comes from its ability of integration within ubiquitous and pervasive computing environments that can be controlled through effective, flexible, and easy-to-use mobile solutions. Indeed, the integration of mobile systems and embedded systems, within Internet-of-Things (IoT) applications, enables the development of powerful devices and applications that contribute to people's well-being.

In this chapter, an abstract IoT system Development Model (IDM) is presented. The model comprises concept refinement pyramids, decision trees, realistic constraint lists, architecture and organization diagrams, communication interfaces, use cases, and menus of analysis metrics and evaluation indicators. The proposed IDM integrates hardware and software sub-systems, and their interfaces. The hardware subsystem includes processors, sensors, interfacing boards, etc. Although the IDM partly adopts sub-models that are common in software engineering; it provides a unique integration and enables effective development, application, presentation, and evaluation. The software subsystem comprises core applications, drivers, databases, etc.; all developed for mobile applications or interfaces. The research objectives of the work presented in this chapter are as follows:

- Develop a model that enables the design, implementation, analysis, and evaluation, and presentation of an IoT system. The proposed model integrates hardware, software, and mobile solutions. The proposed model aims at being clear, concise, visual, easy to use, and effective in application.

- Develop sub-models that can capture the concept, functional requirements, and non-functional requirements of an IoT system. The sub-models are tools that aid the capturing and representation of system functions, technologies, architecture and organization, system interactions.
- Integrate hardware and software systems in development.
- Integrate mobiles solutions in development.
- Integrate main realistic constraints and analysis metrics in development.
- Integrate evaluation indicators in development.
- Target academic users within the context of computer engineering and science.
- Target industrial users.
- Evaluate the model usability and effectiveness in application by developing a variety of IoT systems. All the developed systems adopt the proposed model and a Capstone Design Project (CDP) setup in an undergraduate program on Computer Engineering.

The IDM is used to create and represent several IoT systems. The developed systems include AgriSys, NFC Wallet, and RECON [1-3]. The systems enjoy several creative features including the successful integration of mobile solutions. AgriSys is an agriculture system that deals with desert-specific challenges, such as, dust, infertile soil, wind, low humidity, and the extreme variations in temperatures. NFC Wallet develops a universal card machine that enables the use of highly-secure wearable bracelets to completely replace one's cards and keys in payments and secure access. In addition, RECON is a long-distance delivery and monitoring drone controlled at one's fingertips using a smart device over the Internet. Indeed, all the proposed IoT systems are ubiquitous, they integrate mobile solutions, and contribute to people’s well-being.

This chapter is organized so that the Section 2 presents related work. Section 3 presents the development model. In addition, Section 4 presents sample applications, while Section 5 thoroughly analyzes and evaluates the proposed model. Section 6 concludes the chapter and sets the grounds for future work.

## 2 Related Work

IoT is a collection of various technologies that work together in tandem. Numerous attempts have been made to realize an IoT framework that distills the technological complexities of the IoT into a single multi-layered system architecture. Yet, there is no single consensus on the best development model for IoT. Different reference models have been proposed, for each IoT layer, by different researchers. Whilst a group of researchers consider the development of reference architectures can lead to a faster and exponential increase of IoT-related solutions, other groups believe that the real impact of IoT will be felt in realizing a development model at either the application or the communication level. However, an Integrated Development Environment (IDE) that capture architectural, behavioral and communicational aspect of IoT will ultimately take IoT to the next level. In the following subsections, we explore existing IoT development models, devices, and enabling services.

## *2.1 IoT Development Models*

A complete suite of tools spanning both hardware and software is presented in [4]; namely, Michigan's IoT Toolkit. This toolkit has five components; a gateway, four hardware building blocks, multiple sensor platforms, an indoor localization system, and a software for connecting users and devices. To address the pitfalls of current applications (limited to specific devices, protocols, ...), Michigan's IoT provides a layered interoperable system that offers support for moving single-device, cloud-centric applications and towards richer applications that incorporate multiple data streams, human interaction, cloud processing, location awareness, multiple communication protocols, historical data, access control, and on-demand user interfaces under variable external lighting conditions. In [5], Fuller et al. presented a graphical system design approach that includes hardware abstraction, a heterogeneous multi-processing environment and different models of computation.

The emergence of a common reference model for the IoT domain and the identification of reference architectures can lead to a faster and more focused development of IoT solutions. In [6], an Architecture Reference Model (ARM) is developed. ARM attempts at standardizing the implementation of IoT systems based on an integrated, common approach and an accessible, reference architecture. ARM sustains the interoperability of solutions at the communication level, service level and across various platforms. Also, ARM enables the users to derive use-case and application-specific architectures. In [7], the authors introduced and discussed two reference architectures, the ARM [6] and the architecture proposed by WSO2 [8]. By analyzing the characteristics of these architectures, the authors intend to shed the light on important issues that will take the research on reference architectures to a higher level. The two proposals were analyzed in terms of their support for addressing the main requirements of IoT systems.

In [9], the authors proposed a cloud-based development environment focused on building WSN applications for IoT, called COMFIT. COMFIT enables the users to develop IoT applications with a simple web interface that requires only a browser. The user's computer is not responsible for most of the processing performance since all the compilers and simulators are hosted in the cloud. COMFIT is composed of two modules. The first module is the App Development Module, which is a Model

Driven Architecture infrastructure; it is responsible for providing the application developers with the proper abstraction to support the different phases of application development lifecycle (design, validation, deployment, maintenance, etc.). The second module is the App Management and Execution Module; it is responsible for providing the functionality of uploading the generated code to the remote server hosted in the cloud. In [10], the authors proposed four design patterns (Problem, Context, Motivation Forces, Solution Details) enabling IoT architects to construct edge applications. The defined patterns are reusable, inter-related, well-structured, providing efficient and reliable solutions to recurring problems discovered by IoT system architects.

In [11], a model-driven methodology for the design and development of smart IoT-based systems is presented. Inspired from the human nervous system and cognitive abilities, a set of autonomic and cognitive design patterns is presented, requirements while taking into consideration big data and scalability management, to incrementally refine the system functional and nonfunctional requirements. A cognitive monitoring system for managing the patient health based on heterogeneous wearables is developed to demonstrate the efficiency of the proposed patterns.

In [12], the authors proposed an IoT domain model based on recent available applications from real world. Their model is derived from the extraction of the concepts and the associations that represent the applications in IoT. Concepts can be “Traditional Internet” concepts (computational service, storage service, end user application, etc.) or “Thing” concepts (EOI, resource, action, device, sensor driver, etc.). Associations represent the relationship between the concepts (observe, produce, run-on, host, etc.). Three IoT applications; shared book reviews, smart plants, and room HVAC maintenance; were successfully modeled using the proposed domain model.

## *2.2 IoT Devices and Enabling Services*

IoT devices need mechanisms to connect to the Internet to communicate with other devices and to supporting servers. The communication between IoT devices is mainly wireless because they are generally installed at geographically dispersed locations. The leading communication technologies used in the IoT world are Wi-Fi, Bluetooth, ZigBee, NFC, 2G/3G/4G cellular, and other proprietary protocols for wireless networks. Depending on the application, factors such as range, data requirements, security and power demands and battery life will dictate the choice of one or a combination of technologies.

Today, the most common hardware platforms used by IoT system designers are Arduino, Raspberry Pi, Phidgets, National Instruments, to name but a few. The deciding factor in choosing among these devices can come down to precision, speed, ease of programming. Each has differences in language used, ease of use, cost, and reliability. Arduino is an open-source electronics platform or board and the software used to program it [13]. Some Arduino boards have built-in Wi-Fi and Ethernet capability, plus Linux processors to support more complex functionality. The Arduino boards are programmed through USB-to-serial adapters. The LabVIEW Interface for Arduino (LIFA) Toolkit is a FREE download that allows developers to acquire data from the Arduino microcontroller and process it in the LabVIEW Graphical Programming environment. The advantage of Arduino specifically, and part of what makes it so popular when compared to other development boards, is that there is a standard connector scheme that allows shields to be attached. Shields enable connections to motors, cameras, displays, sensors, et cetera, and are usually necessary in prototypes. The design can be stand-alone, all the code can reside on the device, allowing for a cost-effective and portable design. Arduino makes several different boards, each with different capabilities. Arduino Uno is the simplest and most popular board that contains everything needed to support the microcontroller. LillyPad Arduino is a wearable e-textile technology designed with large connecting pads. It can be sewn to fabric and connected to power supplies, sensors and actuators with conductive thread.

Phidgets are USB sensors and controllers that connect computers to the real world [14]. Phidgets offer a wide range of language support, which includes an API for C#, C++, Java, Python, Visual Basic, Ruby, LabVIEW, Max/MSP, and more. Example codes for Phidgets, and for most products in several languages, are readily available and can easily be adapted for any project. Phidgets are computer-dependent, needing to be connected to a computer to run the application code. Communication with Phidgets can be done in the same language as the source code for the application being developed. This allows users to draw on existing libraries and extensions for any given language without having to develop a bridge between these different environments. Several prototypes have attached Phidgets to a Raspberry Pi, which runs the application, allowing for a fairly cost effective and portable design. The most popular Phidget boards are pH Phidget, Sonar Phidget and Phidget RFID Board USB.

Raspberry Pi is a small-size Linux computer; it's fully functional to the point where it can run desktop applications [15]. Raspberry Pi can be programmed in standard languages like Python and comes with it already installed. Raspberry Pi has multiprocessing capability and can run more than one application at a time; It can easily drive video displays and USB devices, like keyboards and mice, but is harder to wire up to sensors. Debugging is easy on Raspberry Pi, with more advanced tools available and the ability to run Python interactively.

National Instruments' (NIs') products are intended for engineers and scientists needing to measure and control in areas that require a high degree of precision and accuracy [16]. They provide an integrated hardware and software platform with a graphical system design that abstracts complexity. NI offers the RIO family of FPGA-Based deployment targets and the CompactDAQ data acquisition controllers and chassis.

On the software side, several commercial, and open-source, IDEs and services are available for IoT systems. Indeed, the available software solutions and services effectively enable mobile services. NI hardware is programmed using LabVIEW software exclusively. LabVIEW is an IDE that has been designed specifically for engineers and scientists building measurement and control systems. LabVIEW is used to program NI systems, however, currently LabVIEW can support a wide range of devices such as Arduino and Phidgets. Systems created using LabVIEW can be published and hosted on a webserver and can be accessed online using PCs, laptops, tablets, etc.

Many online services are available for supporting the development of IoT systems. Web hosts support users with online systems for storing information, images, videos, or any content accessible via the web. For example, GoDaddy.com is a private web hosting company that provides domain registration, hosting, and other e-business services. Arduino IDE is an online platform that enables the users to write the code, access tutorials, configure boards and share the projects. The code is saved to a Cloud, always backed up and is accessible from any device. Arduino IDE automatically recognizes any Arduino board connected to the PC and configures itself accordingly. Moreover, MobiOne is an advanced cross-platform mobile software development environment for creating both iOS and Android applications, as well as optimized mobile web apps. MobiOne runs on Windows system and provides a drag-n-drop interface for creating multi-page applications. MobiOne programming model is based on the web programming model and uses HTML, CSS and JavaScript for customization [17].

Other services include Amazon Web Services (AWS) IoT; it is a platform that collects and analyzes data from internet-connected devices and sensors and connects that data to AWS cloud applications [18]. AWS Mobile is a platform that provides a range of services that helps in the development of mobile apps. AWS Mobile includes a variety of tools, including tools to track application analytics, manage end-user access and storage, set up push notifications, deliver content, and build back-end services. Furthermore, Google Cloud IoT is a set of fully managed and integrated services (Cloud IoT Core, Google BigQuery, Cloud Machine Learning Engine, Google Data Studio) that allow to easily and securely connect, manage, and ingest IoT data from distributed devices at a large scale, process and analyze data in real time, implement operational changes and take actions as needed [19].

## 3 The Internet-of-Things Development Model

The proposed IoT Development Model (IDM) enables system development by providing sub-models that can capture the concept, functional requirements, and non-functional requirements. Moreover, IDM sub-models are tools that aid the capturing and representation of system functions, technologies, architecture and organization, system interactions, main realistic constraints, and analysis metrics and evaluation indicators. IDM comprises Concept Refinement Pyramid (CRP), Decision Trees (DTs), Architecture and Organization Diagram (AOD), Communication Interfaces (CIs), Use Case Diagrams (UCDs), Realistic Constraint List (RCL), Menu of Analysis Metrics (MAMs), and Evaluation Indicators Graph (EIG). The CRP is a pyramid that refines the system top-level concept to Functions and Technologies. Usually, the CRP is of three levels. Each level is described with short statements. The top of the CRP is the concept that must be defined using a title. The purpose of the CRP is to provide an illustration of the system concept, design, and implementation highlights. The highlights of the Functions and Technologies levels address issues that comprise system currency, novelty, design functions, implementation technologies, adherence to professional ethics, and attention to global, economic, environmental, and societal factors.

DTs, UCDs, and AODs are refinements of the functional requirements of the system. A DT is a multi-leaf-node tree that illustrates alternate design or implementation options at the Function or Technology levels. DTs show design and implementation alternatives and options. One of the DT nodes is selected as the designer choice. Evaluation characteristics are included in the tree nodes to justify the selection. Moreover, the IDM adopts the UCDs of UML to depict the system interactions. UCDs include actors that interact with the different system functionalities, such as managing, using, extending and including. Besides, UCDs specify the events of a system and their flows.

The AOD is the core of the IDM; it depicts the architecture and organizational structure of the system. Moreover, it shows the communication of the different technologies, subsystems, and platforms employed. The presented AOD can adopt three variations of the Communication Interface (CI). The first CI model relies on services hosted by a third-party Internet provider. Here, the interfacing device and the user fully communicate through the third-party providers, such as, a webserver, database server, application server, etc. The second CI model comprises a server-based interface with a dedicated Internet Protocol (IP) address. In this model, the user communicates with the server hosted on the processing system. The third CI model enables a direct communication between the user and the processing system when the user and the device are at the same location. For the three proposed variations, the user communicates with the system using a mobile solution that can be an application or a web-interface using a browser.

RCL, MAMs, and EIG are refinements of the non-functional requirements of the system. An RCL is a list that highlights the realistic constraints of the system. Realistic constraints can be tight; they are, usually, critical requirements during system development. The clear identification of the system constraints; using the RCL; highlights their importance, enables their adequate satisfaction, and present them in a self-contained diagram. MAMs are of two types, namely, general and application-specific. General metrics are common to all systems developed based on the IDM, however, application-specific metrics are concerned only with the developed system. MAMs enable the evaluation of the robustness, thoroughness, and adequacy of the system and its chosen metrics.

The Evaluation Indicator (EI) is a combination of a set of Key Indicators (KIs) that aid the evaluation of the developed system based mostly on IDM components. The identification and clear understanding of the evaluation KIs enables the early integration of their requirements in the design and implementation stages. Usually such KIs are specified in project evaluation criteria whether in academia, industry, competitions, etc. [20].

The EIG adopts a holistic approach in evaluation. The evaluation scale of each KI is holistic and range from 1 to 5, where 5 is the highest score. The EI is the arithmetic average score of all KIs. In addition, analytic approaches can be adopted with minimal tuning to the EIG scale [20, 21]. Table 1 presents the EI key indicators and their mapping onto the IDM components. A typical development procedure using the IDM is identify the system concept; review the EIG KIs; survey the literature to identify and properly acknowledge closely-related work; identify the main system features; identify global, economic, environmental, and societal factors; develop the CRP; develop the RCL; use DTs to identify functions and implementation technology; develop AOD, UCD, and MAM; implement the system; analyze the system; evaluate the system; and then draw the Final IDM Chart. A sample generic IDM chart is shown in Figure 1.

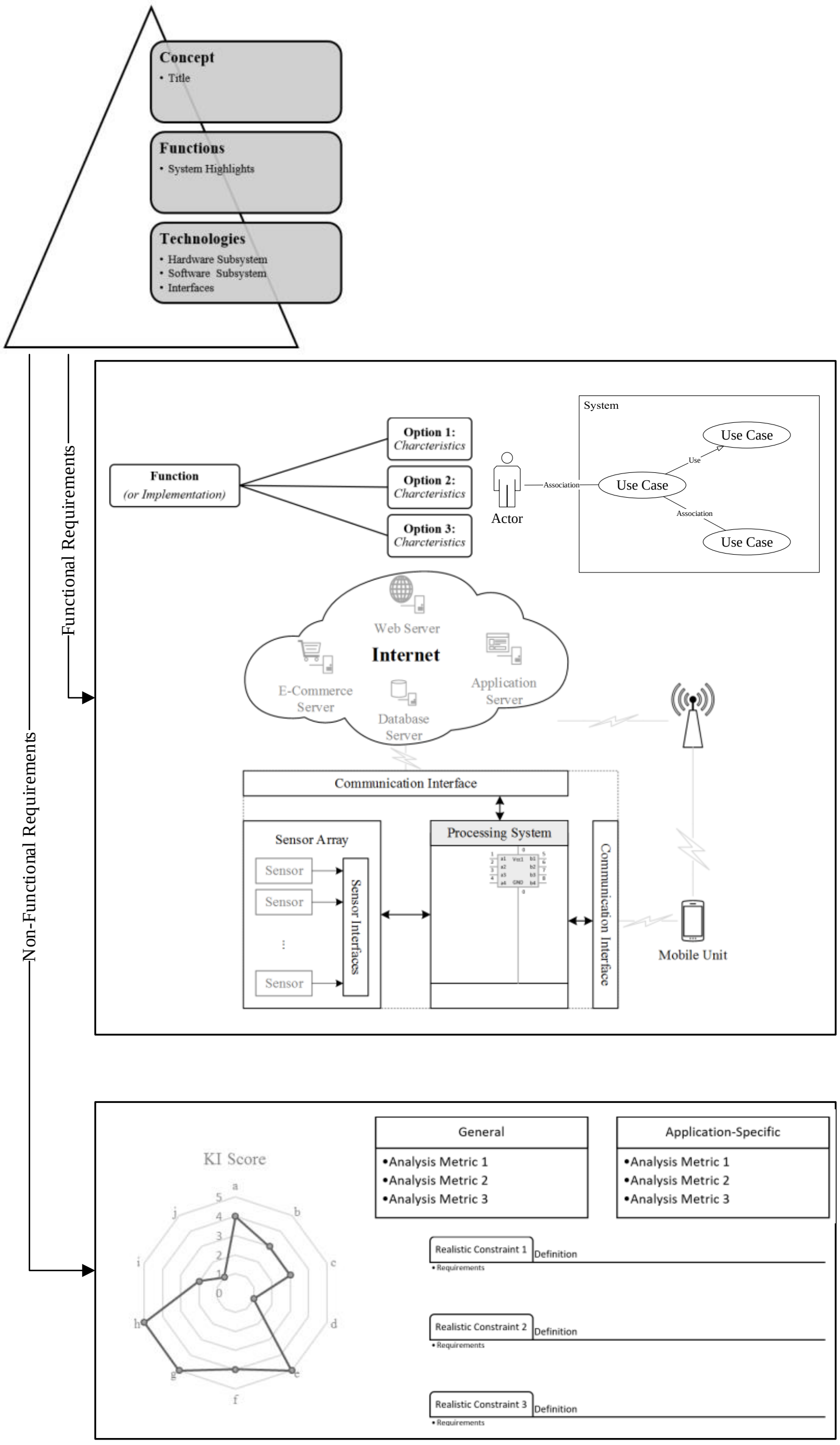

**Fig. 1. A sample generic IDM Chart that shows all the proposed models. the chart includes generic CRP, AOD, UCD, DT, MAM, RCL, and an assumed EIG.**

**Table 1. Evaluation Indicators and their mapping onto IDM components.**

| EI | Key Indicators [IDM] |
|---|---|
| **a.** | Currency and novelty **[CRP]** |
| **b.** | Design **[CRP, AOD, UCD, DTs]** |
| **c.** | Alternate design options **[DTs]** |
| **d.** | Implementation technology **[CRP, AOD, UCD, DTs]** |
| **e.** | Alternate implementation technologies **[DTs]** |
| **f.** | Identification, mastering, and use of hardware/software tools **[DTs]** |
| **g.** | Consideration and adequacy of standards and realistic constraints **[RCL]** |
| **h.** | Robustness, thoroughness, and adequacy of analysis metrics **[MAMs]** |
| **i.** | Adherence to professional practice and code of ethics **[CRP]** |
| **j.** | Attention to global, economic, environmental, and societal factors **[CRP]** |

# 4 Sample Model Application Systems

To present the proposed model, three system applications are presented in the following subsections. The presented systems are AgriSys, NFC Wallet, and RECON. The selected systems belong to different areas of application, namely, agriculture, wearable security systems, and unmanned aerial vehicles. The proposed systems follow the same development patterned that can be captured using the IDM.

## *4.1 AgriSys: A Smart and Ubiquitous Controlled-Environment Agriculture System*

AgriSys is a smart Agriculture System that can analyze an agriculture environment and intervene to maintain its adequacy. The system deals with general agriculture challenges, such as, temperature, humidity, pH, and nutrient support. In addition, the system deals with desert-specific challenges, such as, dust, infertile sandy soil, constant wind, very low humidity, and the extreme variations in diurnal and seasonal temperatures. For a reduced controller complexity, the adoption of fuzzy control is considered [1]. The following requirements define the functions of AgriSys:

- Sense temperature
- Sense humidity
- Sense soil moisture
- Sense soil temperature
- Sense the light
- Sense the pH
- Response to sensor readings to control irrigation
- Response to sensor readings to control a fan
- Response to sensor readings to control a shutter

AgriSys is designed to offer low cost, efficient and accurate control options. AgriSys is implemented to help saving and reducing the amounts of used water and energy. Indeed, forming cross-disciplinary teams of agriculture specialists and engineer to enable the modernization of agricultural procedures. From an economic point of view, agriculture is the source of livelihood for more than half of world's population; AgriSys increases productivity of farmers at an affordable fixed cost. AgriSys is novel in the sense that it targets extreme weather conditions using a fuzzy controller and enable remote control over the Internet.

AgriSys is implemented using Phidgets interface8/8/8 which connects to a variety of sensors. Moreover, AgriSys system is used to control the temperature, waterfall and sunlight reaching the plants which is essential for greenhouses system. Phidgets sensors are used for the detection of the required environmental variables such as the humidity, temperature, pH and water level, water and soil temperature, soil moisture, and light. The system also includes a pH or Oxidation-Reduction Potential (ORP) Adapter Interfaces. The interface is to a pH or ORP glass electrode through another connector, the BNC connector, and gives it the data it needs to an input on the Phidgets Interface board. Moreover, the system has a Type-K Stainless Steel Thermocouple with hot and cold junctions. The hot junction is the end that is inserted in the environment of interest, and the cold one is the one used to obtain the readings from the sensor. A pump is needed in both the irrigation and the cooling system to sprinkles water depending on the soil temperature, plant humidity, and the soil moisture sensors.

AgriSys is developed under LabVIEW that comprises a large set of tools for the acquisition, monitoring, analysis, and data recording as well as tools to help debugging code. The complexity and number of used components are reduced with the use of a fuzzy inference system. Using the fuzzy logic library under LabVIEW, the system includes five inputs and different membership functions that also include outputs. AgriSys is deployed on a webserver to enable distant access and monitoring using tablets, computers, smartphones, etc.

In all, AgriSys provides increased productivity, enhanced safety, instant interventions, and an advanced life style. The system is ubiquitous as it enables distant access. Using a fuzzy controller helped mimicking the behavior of a human operator and added another dimension to the novelty of AgriSys. AgriSys is an addition to the current state-of-art IoT systems.

The IDM components of AgriSys are presented in Figures 2 through 9. In Figure 2, the CRP of AgriSys is presented. In Figure 3, a DT is used to decide on what type of system-controller to use, while the DT in Figure 4 enables the decision of which hardware and software integration to adopt. Figure 5 shows the AOD of AgriSys. The system has a variety of sensors and actuators. The NI Compact-DAQ system is interfaced through a computer or a tablet using a USB port. The computer is configured as a server that runs a published LabVIEW control page with a dedicated IP. The system can be remotely controlled using regular computers or mobile systems through web-browsers. The recommended mobile system is a tablet running an MS Windows operating system; this is a requirement to run LabVIEW plugins. A user can control the system from within a Wi-Fi network or using the Internet. Figure 6 shows the UCD of AgriSys. In Figure 7, three realistic constraints are listed, besides, the corresponding design decision. Figure 8 lists the general and application specific measurement and analysis metrics. Indeed, applying IDM on AgriSys development allows the refinement of the functions, technologies, decisions realistic constraints identified in the CRP, DTs, and RCL; and captured in the AOD. Figure 9 shows the EIG of evaluating AgriSys. The evaluation is done by three professionals in the field.

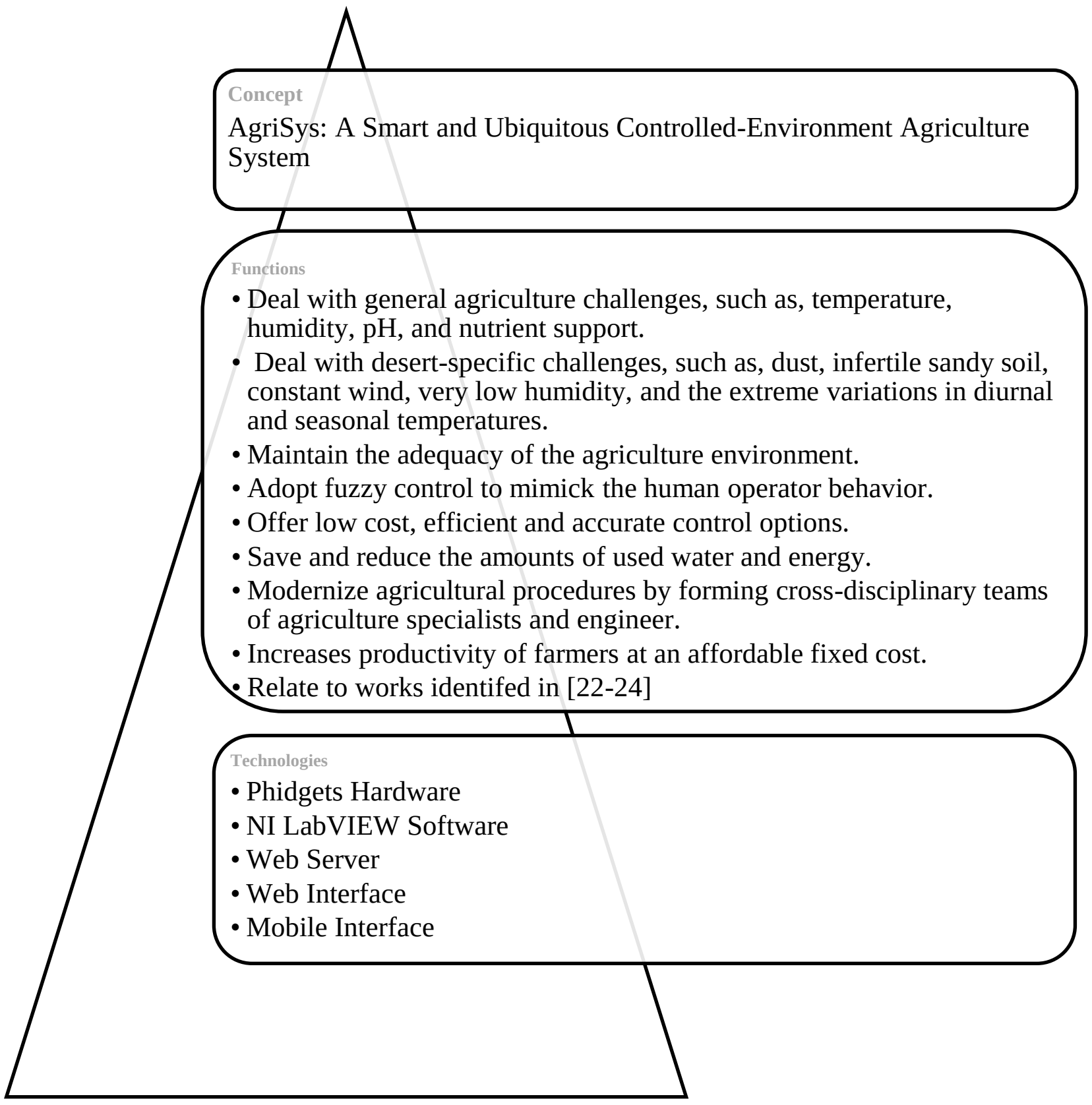

**Fig. 2. The CRP of AgriSys.**

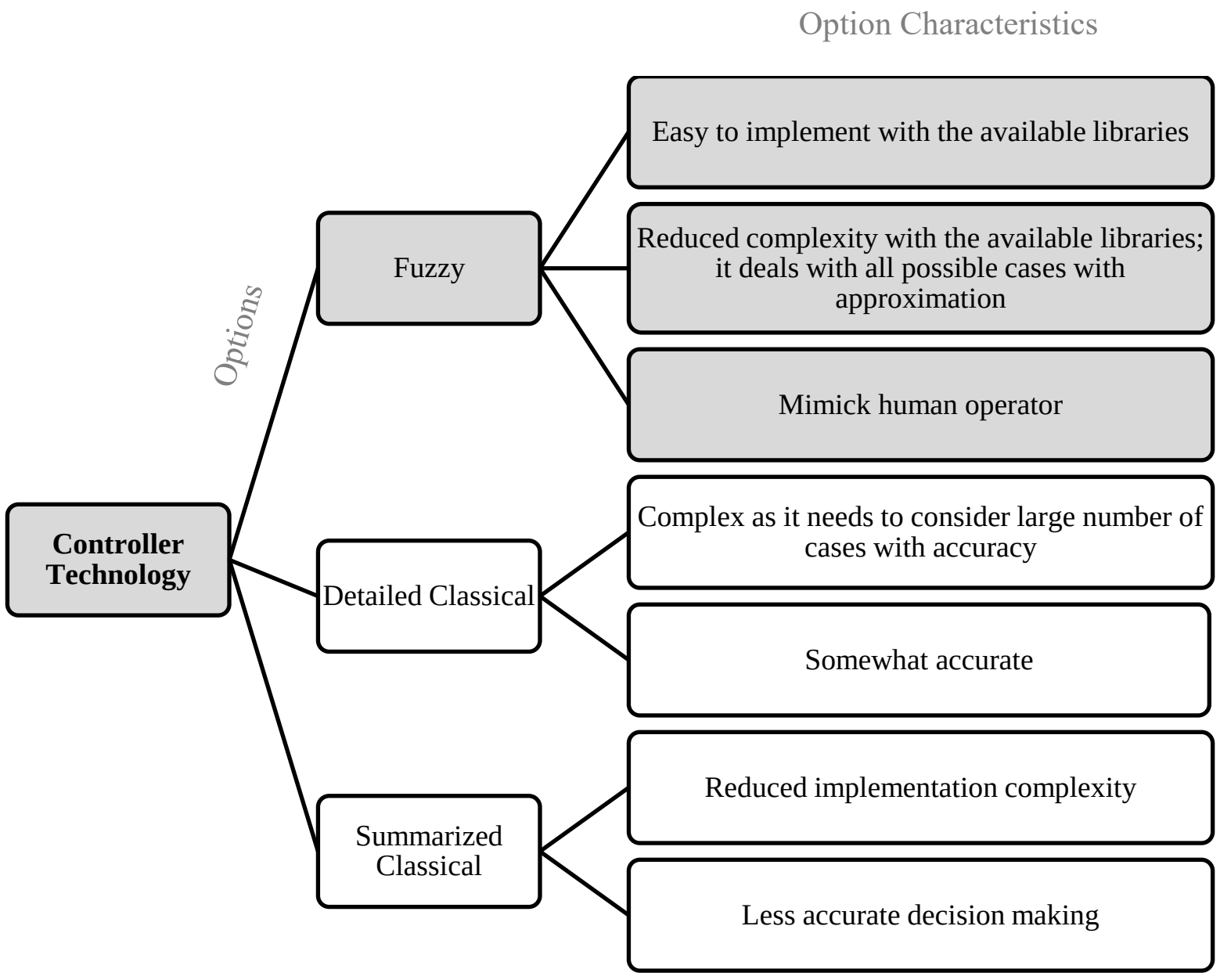

**Fig. 3. The AgriSys DT of Controller Technology. The decision branch is highlighted.**

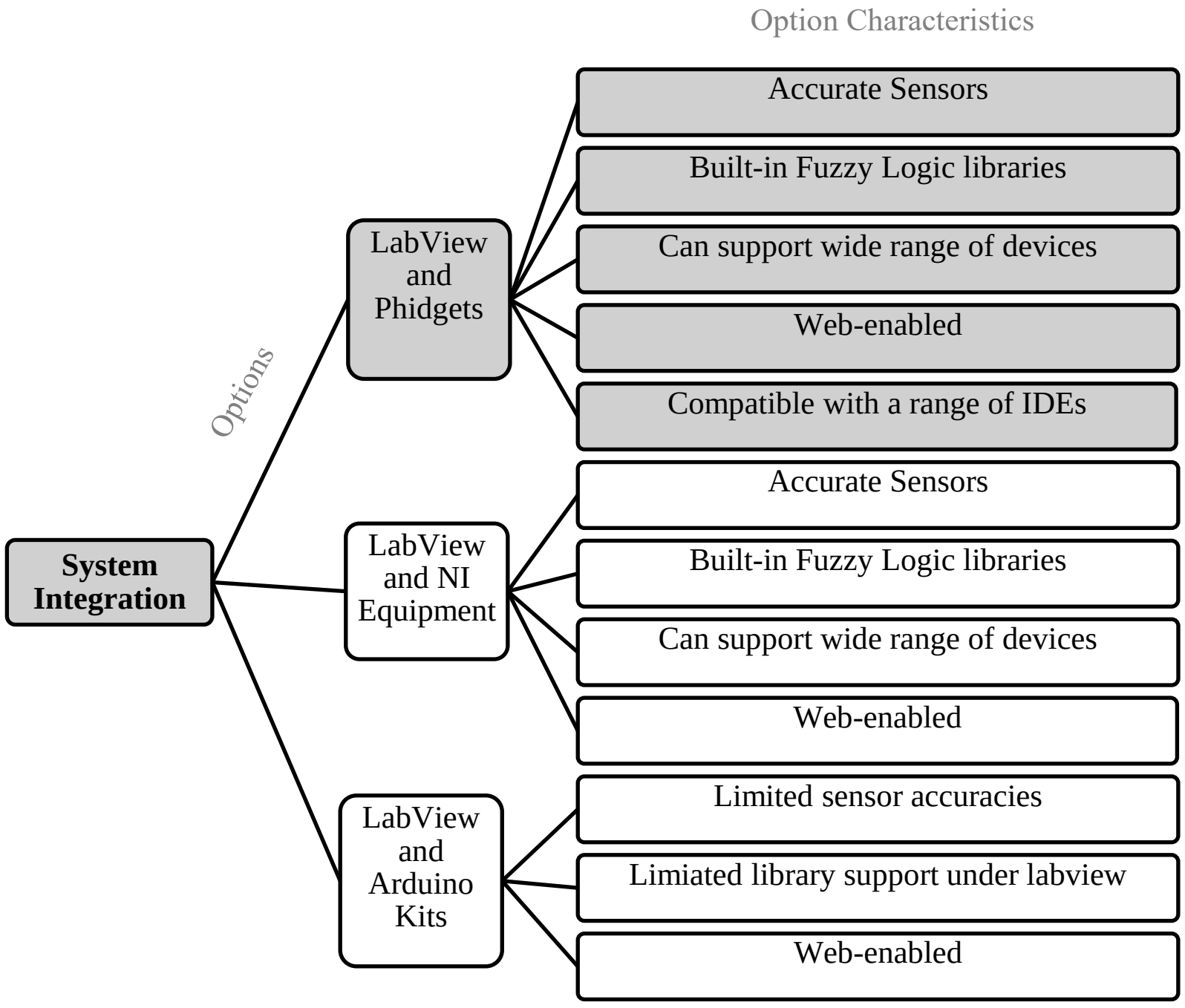

**Fig. 4. The AgriSys DT of hardware and software system integration. The decision branch is highlighted.**

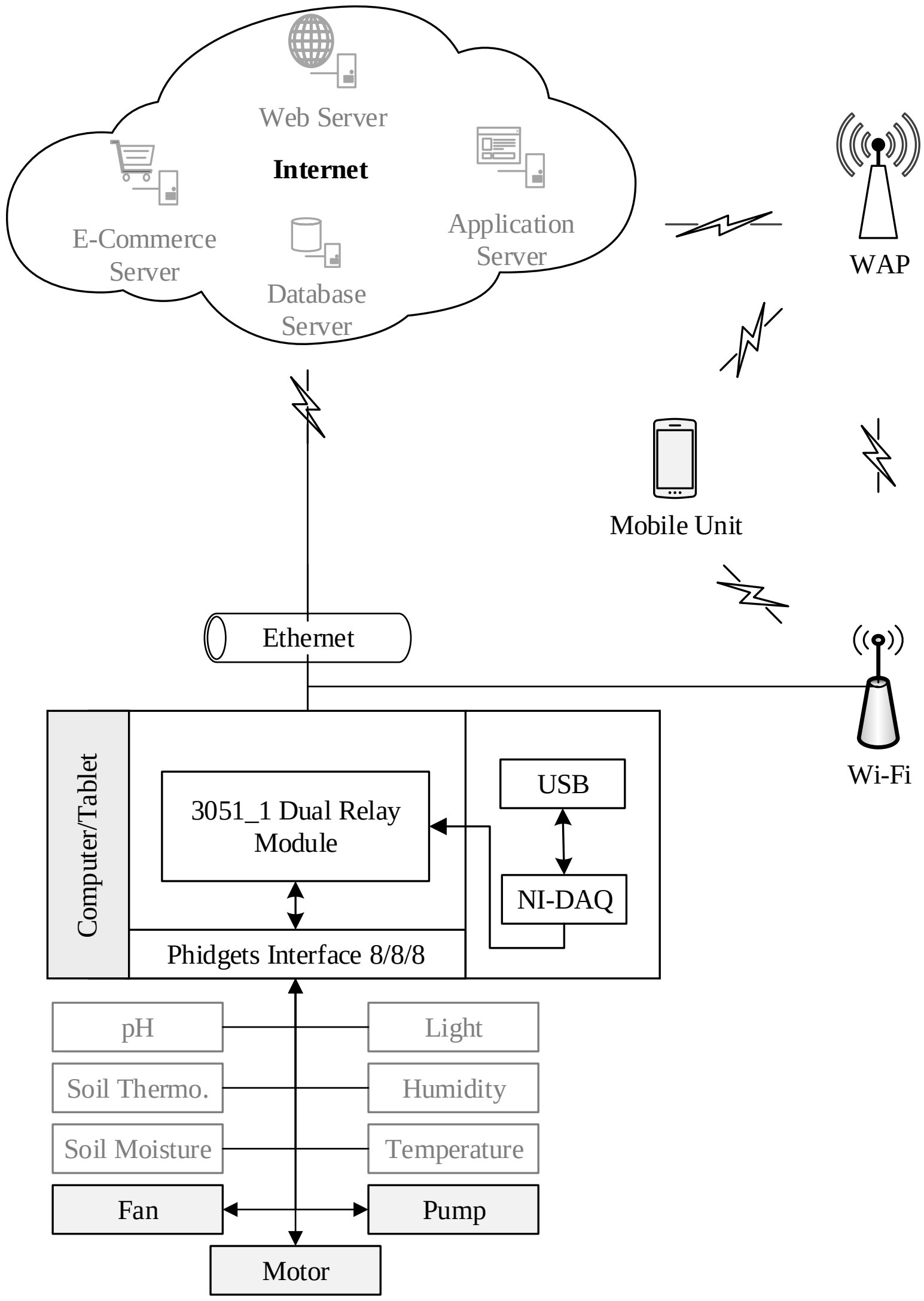

**Fig. 5. AgriSys AOD.**

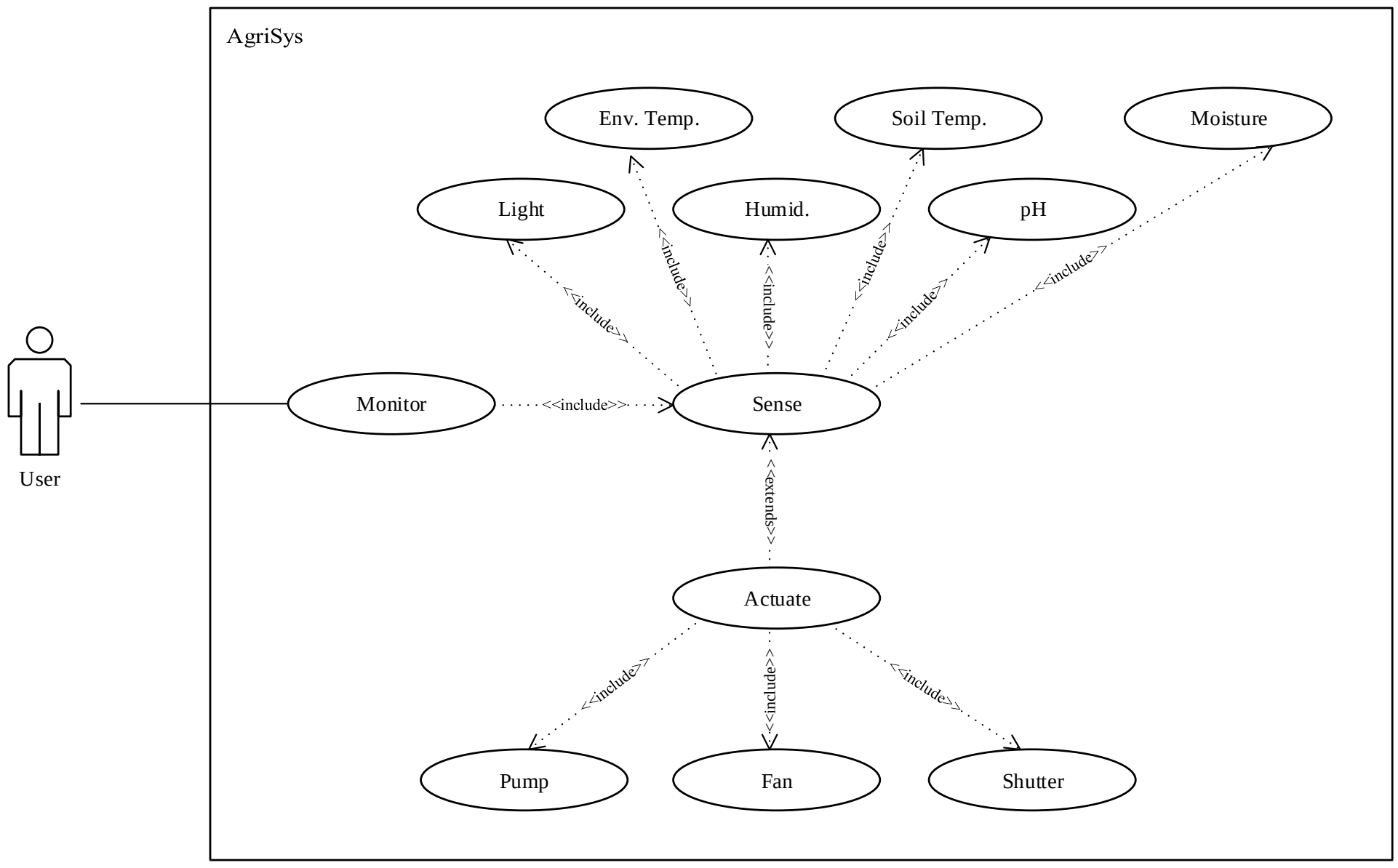

**Fig. 6. AgriSys UCD.**

**Extreme Weather Conditions** — Stormy weather in deserts affect temperature, humidity, soil moisture readings, etc.

- Rely on different sensor readings to shut the plant house.
- Possible system expansions include adding distant and close dust sensors, and nutrition sensors

**Curved Roof Shape** — Parts of the plant can suffer greatly from insufficient light due to a curved roof shape

- Design a shutter to regulate the amount of sunlight incident on the plant regardless of the roof shape

**Low Market Awareness** — Farmers usually rely on traditional ways to increase production

- Raise awareness among farmers to show the importance of smart farming applications

**Fig. 7. AgriSys RCL.**

**General**

- Modernity
- Maintainabiliy
- Usability
- User-Interface Adequecy
- Flexibility
- Scalability
- Portability
- Novelty
- Reliability
- Availability
- Dependebility

**Application-Specific**

- Productivity
- Safety
- Procedure Complexity
- Efficiency
- Affordability
- Accuracy
- Resoponse Time
- Measurement Capability
- Actuation Capability
- Power Consumption

**Fig. 8. AgriSys MAM.**

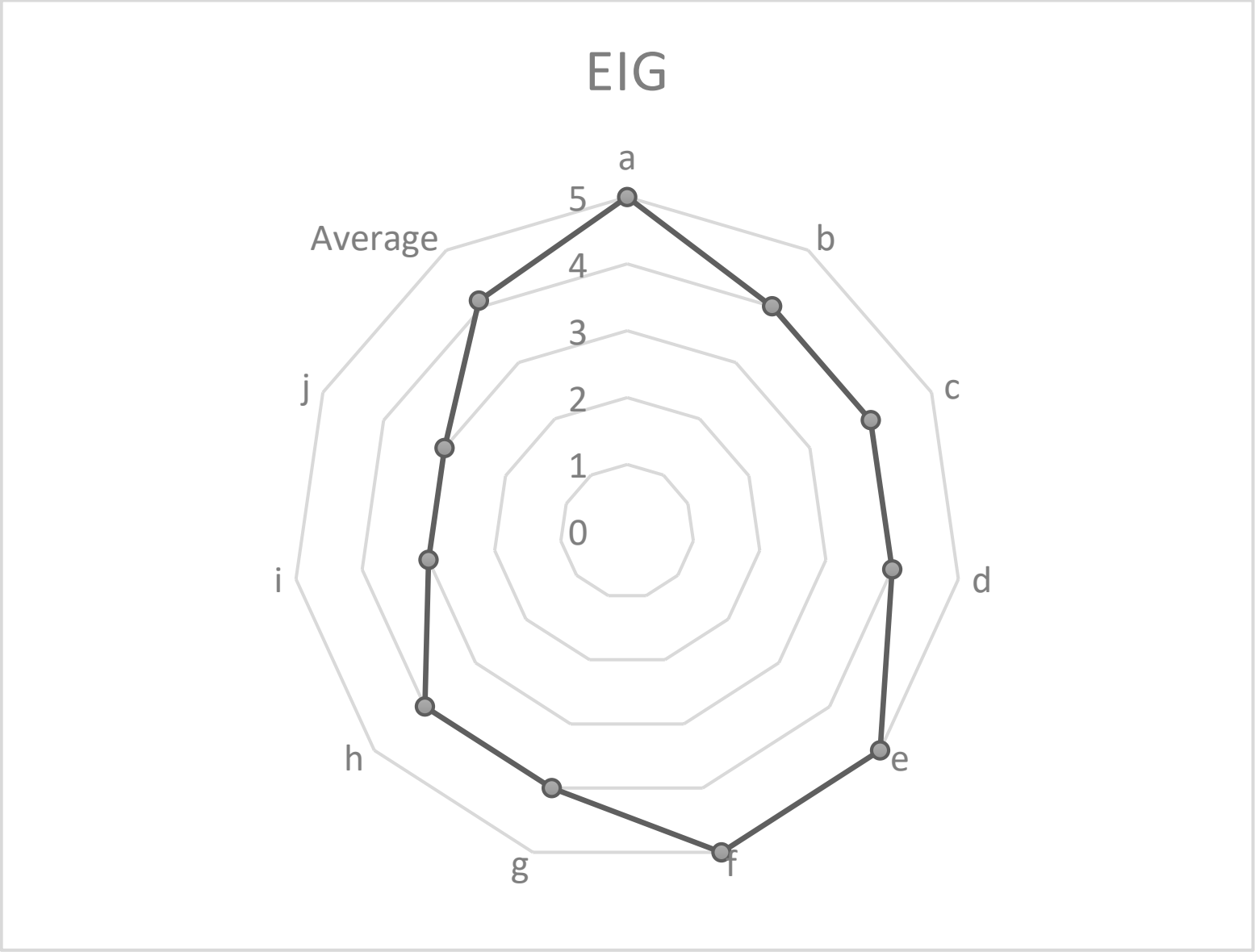

**Fig. 9. AgriSys EIG showing the scores per KIs (a) through (j).**

### 4.2 NFC Wallet

NFC Wallet is a ubiquitous and secure system that enables the use of a single customizable and wearable device to replace traditional smart cards [2]. The proposed system is of three main parts, namely, a wearable device, a control web-enabled phone application and a server, and a reader hardware device. The application on the phone is connected to a webserver through which a multitude of tasks can be accomplished. For instance, one can have his/her credit card, debit card, gym card, office card, and university ID card all stored on one account that has been created on the application. This information is then stored and updated on the webserver and it is identified by a unique ID on the wearable technology. The wearable technology is a ring with an embedded NFC tag. The presented system enjoys several characteristics including its convenient use, secrecy, robustness, reliability, scalability, to name a few.

The purpose of the reader is to receive the information from the wearable technology, and then use it to connect to the webserver and provide access. The reader checks for the account in the webserver's database and enables it depending on its availability. The webserver is supported by a MySQL database. The MySQL database is used to hold users' information, their account information, etc. The database gets all the information from a PHP file, which acts as a mediator.

NFC Wallet is as secured using the AES cryptographic algorithm in the PHP communication layer to ensure the privacy of the data and protect the users' personal information. Two firewall demilitarized zones are created to protect the main server as well as any payment gateways.

The reader contains the following components: Arduino Mega, Ethernet/Wireless Shields, NFC Read/Write Device (RWD), Thermal Printer, TFT Touch Screen, and Electric Door Lock. The reader supports Ethernet or Wi-Fi connections. The AOD of NFC Wallet is depicted in Figure 10. Figure 11 shows the UCD of the functional requirements of the web and mobile application GUI. The main two actors are the User and the Admin. Some of these functions are log in, sign up, and manage accounts by adding or removing them. The UCD also shows some of the exceptions that can be handled as extensions to the use cases; such errors include invalid emails, UIDs or e-mail addresses that have already been registered, etc.

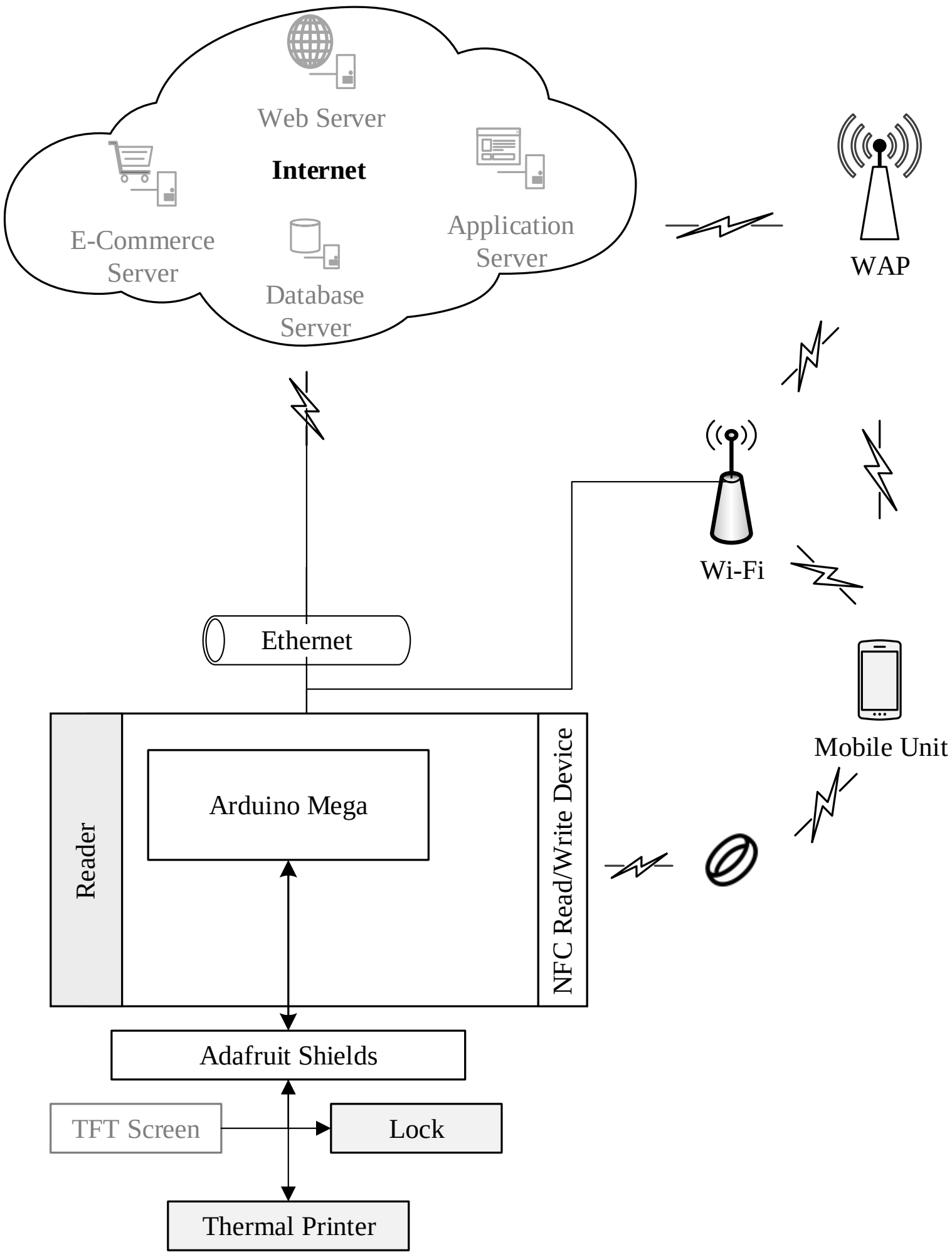

**Fig. 10. AOD of NFC Wallet**.

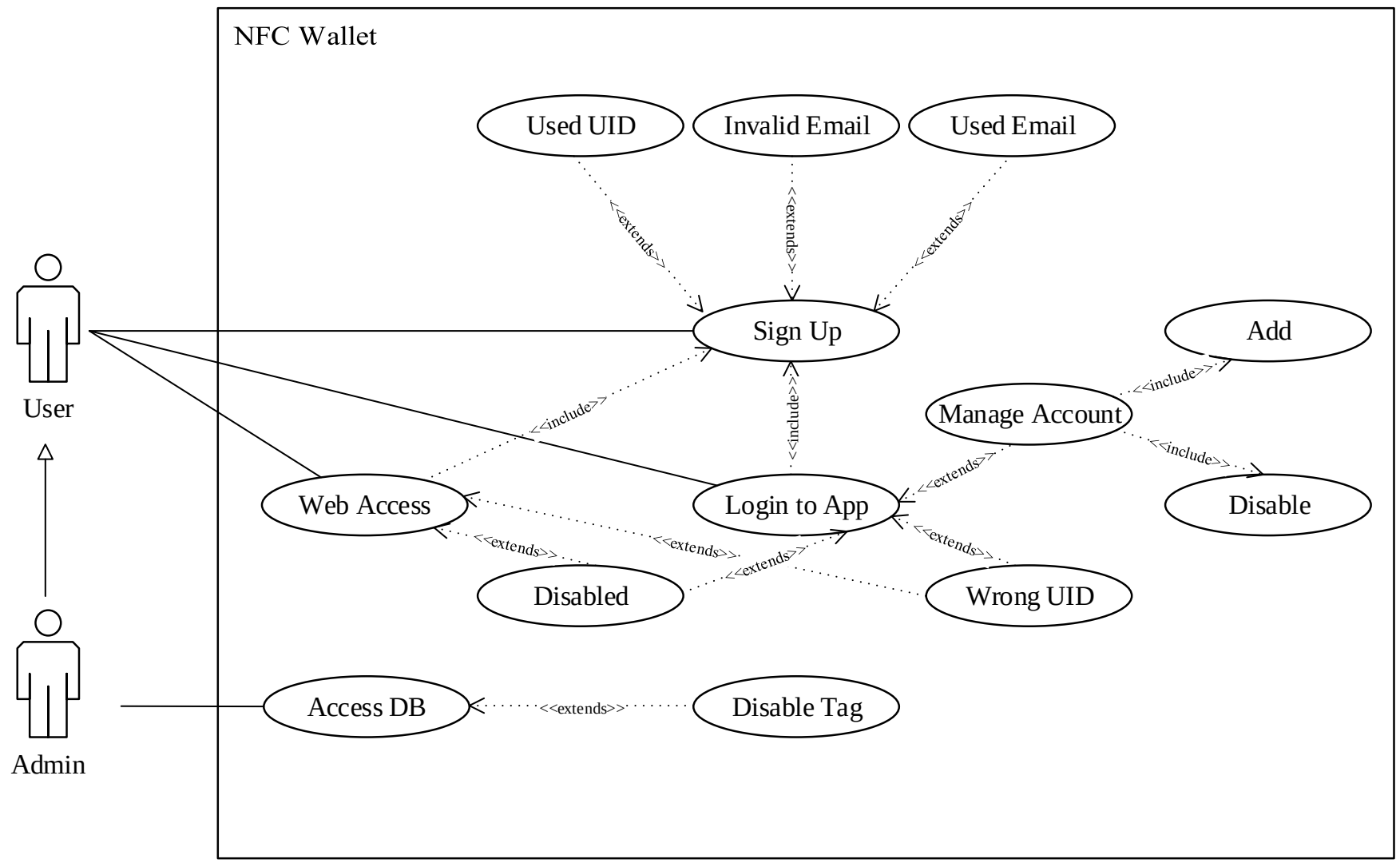

**Fig. 11. Web and mobile application access UCD of NFC Wallet**.

## *4.3 RECON*

RECON is an aerial monitoring system that can yield detailed information to help traffic planners, safety managers, and commuters [3]. RECON provides a "bird's-eye view" for traffic surveillance, road conditions, and emergency response to improve overall quality of transportation. RECON is wired with technology features that monitor the safety, security and activities of their residents to improve overall quality of transportation, search for missing people, and increase independence in technology (search and rescue operations). It can transport goods using various means based on the configuration of the RECON itself. The user can either program a flight path into RECON using a command prompt interface or control the aircraft directly via PC.

RECON system is comprised of a navigation system, camera, speaker, microphone, sensor areas, delivery arm, and centralized control system, and soft interface. RECON is developed under LabVIEW. RECON user can interface with the drone using LabVIEW via a PC, laptop, smart phone or tablet using standard Wi-Fi networks. A critical part in RECON is replacing a potentiometer with LEDs and photoresistors to create the same effect by changing the voltage value passing through the LED. Besides, an NI CompactDAQ controls the output voltage, which drives the LED. RECON is deployed on a webserver to enable distant access and monitoring using tablets, computers, smartphones, etc. The computer is configured as a server that runs a published LabVIEW control page with a dedicated IP. The system can be remotely controlled using regular computers or mobile systems through web-browsers. The recommended mobile system is a tablet running an MS Windows operating system; this is a requirement to run LabVIEW plugins. A user can control the system from within a Wi-Fi network or using the Internet.

In all, RECON is prepared to intervene by communicating, delivering, and assisting according to its capabilities. RECON touches at several aspects, such as security, by using surveillance system and communication system; safety, by using a non-fossil fuel and gyroscope system; advanced lifestyle, by using remote control, and real-time. In addition, RECON ensures having good stability performance while flying and to get real time flight telemetry. RECON is capable of self-stabilization, semi-agile flights, and includes a framework for assorted add-ons, such as, attached cameras or cargo winches. Figures 12 and 13 present the CRP and AOD of RECON.

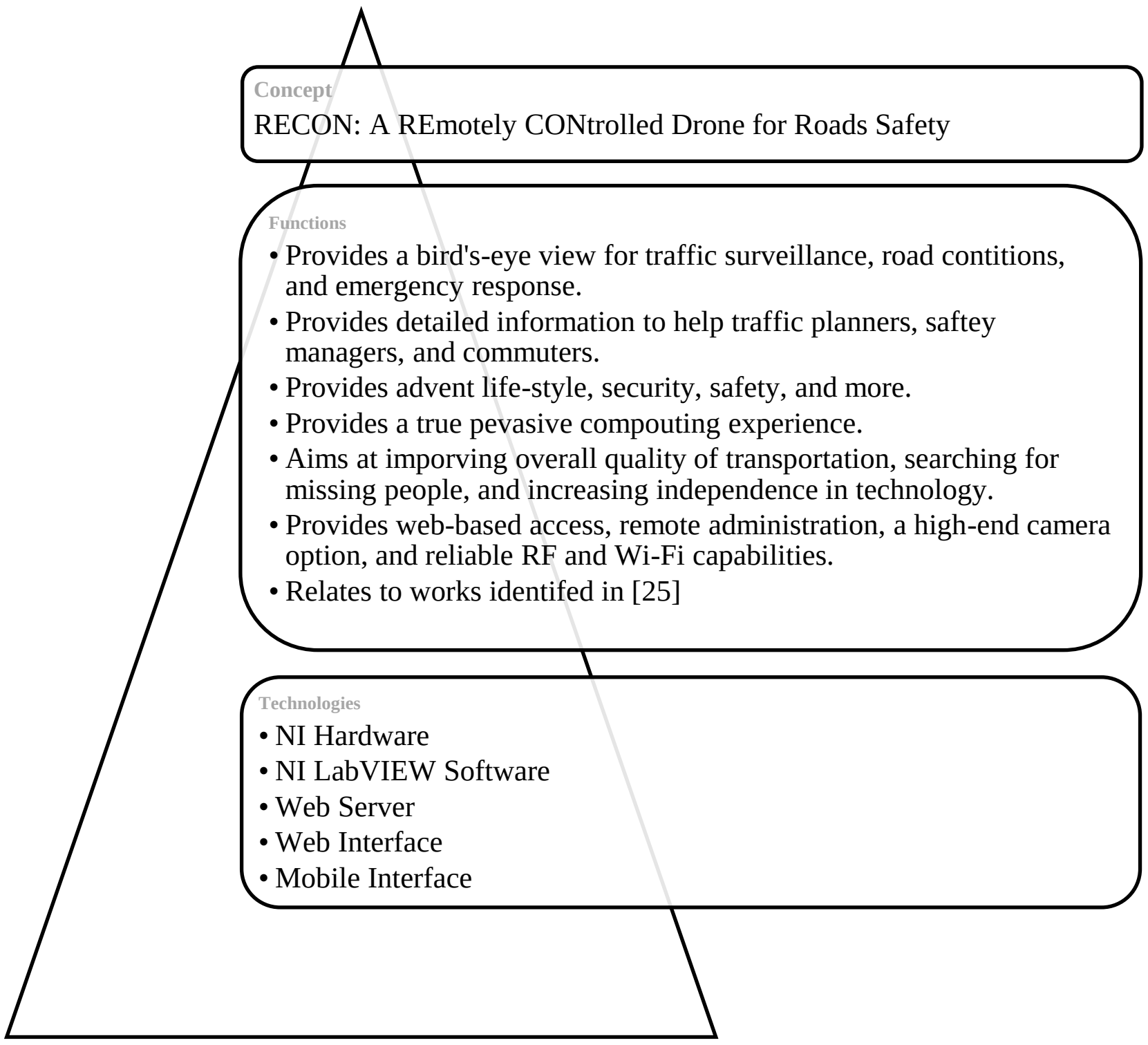

**Fig. 12. CRP of RECON.**

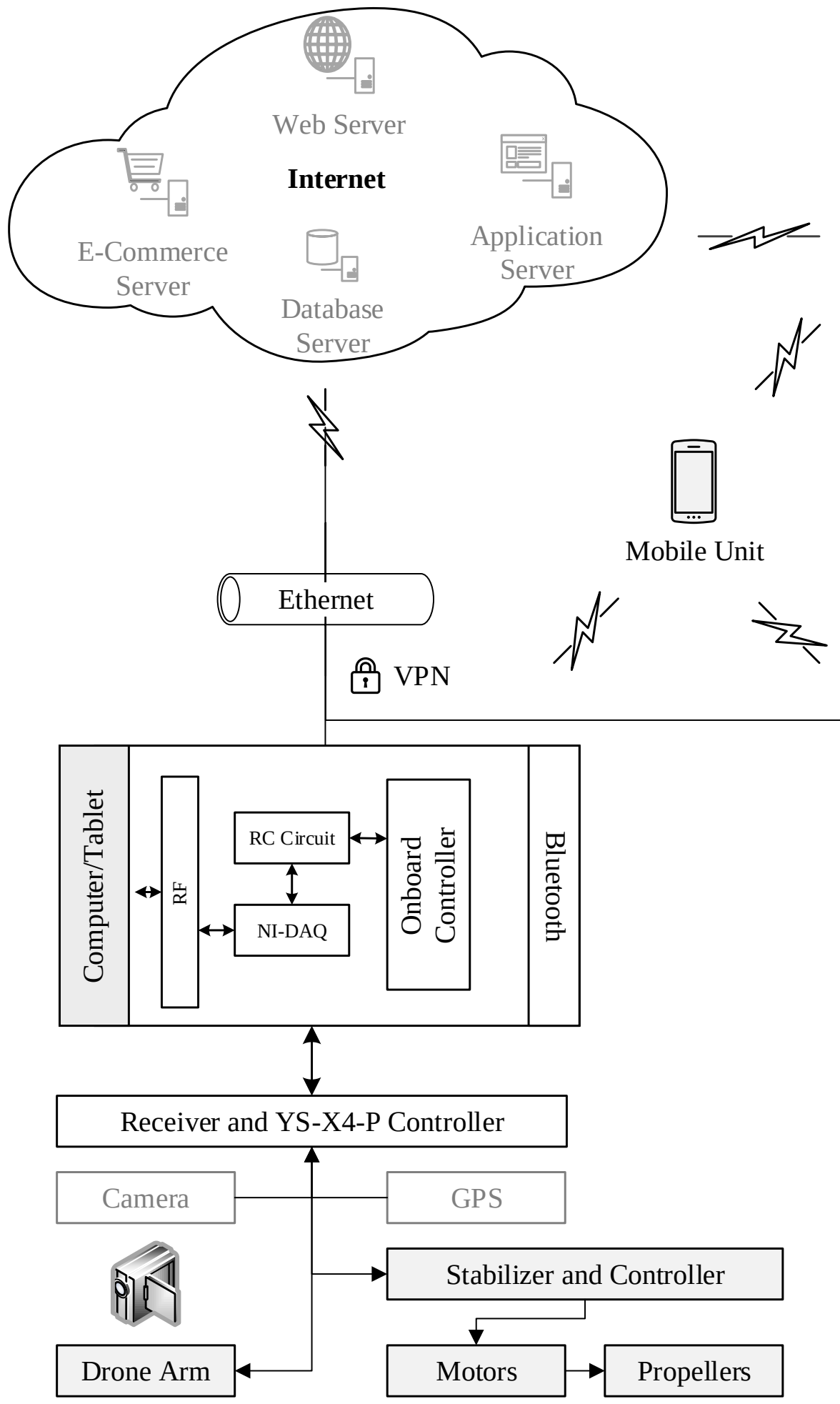

**Fig. 13. AOD of RECON.**

## *4.4 Integration of Mobile Solutions*

The presented applications benefit from the three Communication Interface (CI) patterns of the IDM. All applications benefit from the first CI model, where they rely on services hosted by a third-party Internet provider. Here, the interfacing device and the user fully communicate through the third-party provider, such as, a webserver, database server, application server, etc. RECON adopts the second CI model that comprise a server-based interface with a dedicated Internet Protocol (IP) address. All the presented systems benefit from the third model that enables a direct communication between the user and the processing system when the user and the device are at the same location. Indeed, in all the presented applications, the user communicates with the system using a mobile solution that can be an application or a web-interface using a browser. Figures 14 and 15 show snapshots of the mobile user interfaces of RECON and NFC Wallet.

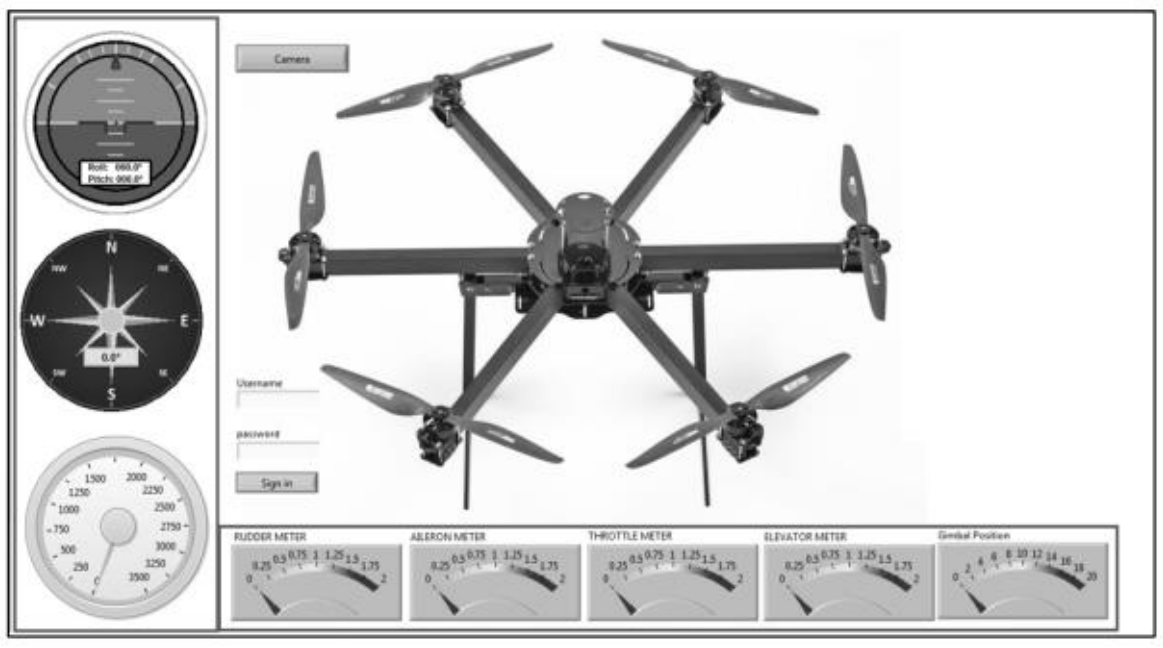

**Fig. 14. The GUI of RECON and as developed under LabVIEW and published on a web-server.**

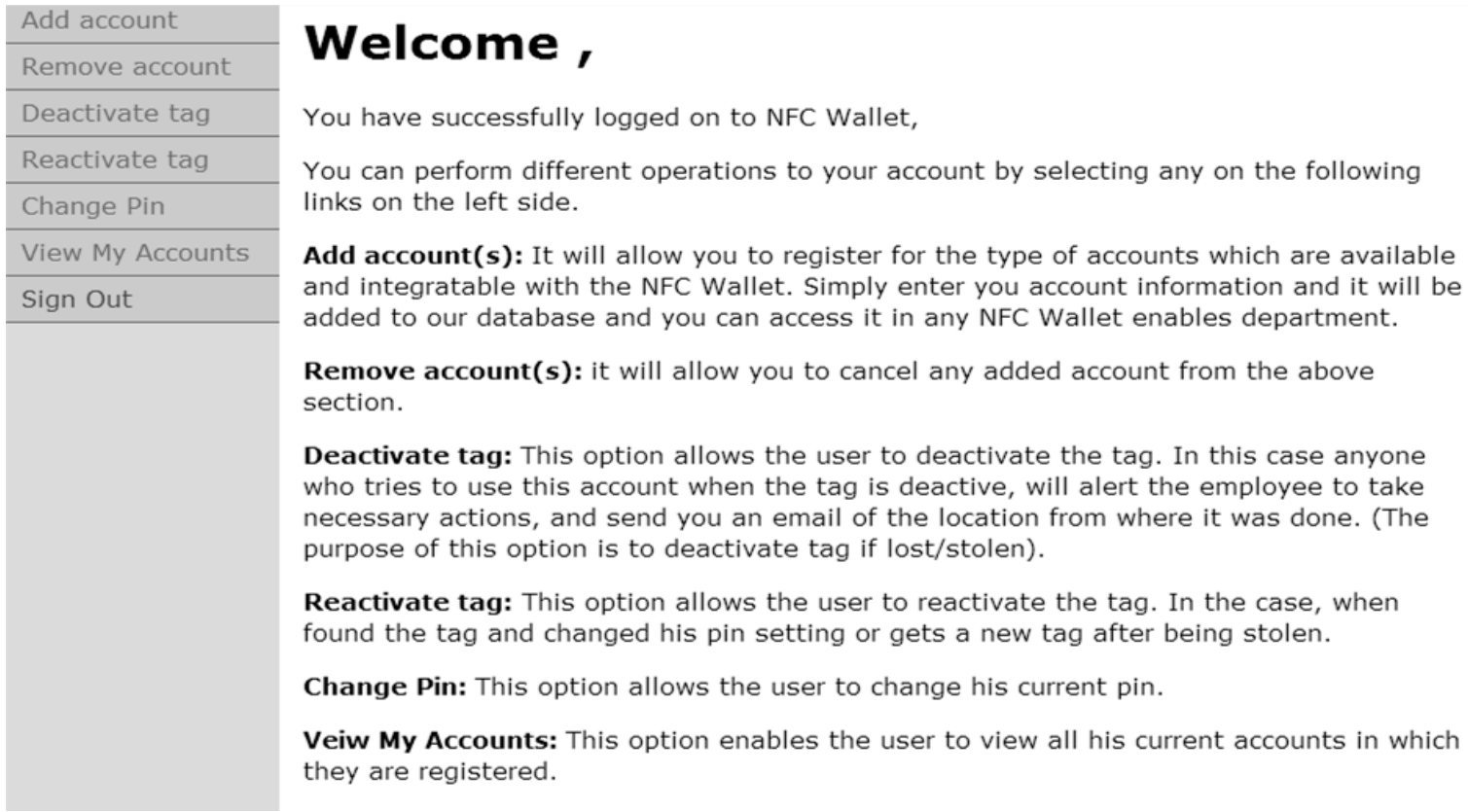

**Fig. 15. The GUI of NFC Wallet mobile-phone compatible webpage.**

# 5 Analysis and Evaluation

The evaluation of the proposed IDM starts by considering the aims, achievements, and advantages of the proposed model. In addition, the model application within higher education is investigated. Moreover, the model effectiveness is evaluated based-on a small-scale empirical study. Finally, the section is concluded by presenting the limitations, challenges, and the identified opportunities for future work.

## *5.1 General Evaluation*

The proposed IDM is a model that enables the design, implementation, analysis, evaluation, and presentation of an IoT system. IDM aims at providing an easy-to-use, clear, systematic and informal, and effective design tool. The focus of the proposed model is on capturing the system concept and its development through design and implementation. In addition, the model aims at enabling the reasoning about various decisions taken during the development process. The IDM provides an abstract and rich view of the domain; such a view can help people new to the field with understanding the particularities and intricacies of IoT. In addition, the IDM aids in identifying independent, reusable, and off-the-shelf building blocks for IoT systems. The model enables the addressing of realistic constraints and evaluation indicators early in the design process. The model is visual and suitable for presentations at various development stages. The high-level view of IDM is of an added educational value. The proposed model has proven to be useful and highly effective in application.

Several returns are noted for the proposed IDM. Based on the system design team opinions, in a focus group setup, the CRP is found to be of great assistance at the early stage of the project. The CRP can facilitate capturing the candidate project ideas descriptions; comparisons among possible alternatives; and with the assistance of the EI Indicators, the selection of one final project idea. The two types of the proposed DTs, namely the Function and Technology, help in focusing on a few alternatives while clearly specifying the advantages of the specified choices. The AOD template is a reusable design pattern that can be adopted to tailor a context-specific instance. At this point, the UCD assists in specifying the desired system interactions. The RCL, MAMs, and EIG are refinements of the non-functional requirements of the system; they allow for the reasoning about, and integration of, such requirements early in the design process. In all, the systems that adopted IDM enjoy the following benefits:

- Facilitated development.

- Clear plan setups.
- Well-defined development steps.
- Increased design productivity.
- Reuse of off-the-shelf design models.
- Concise and visual presentation.

The systems that adopted the IDM possess the following characteristics:

- Pervasive, ubiquitous, and Internet-enabled.
- Modern.
- Have one or more novelty aspects.
- Based-on well-selected design options.
- Based-on well-selected implementation options.
- Demonstrate mastering and use of hardware and software tools.
- Based-on robust, thorough, and reliable analysis and evaluation.
- Adhere to professional practice and codes of ethics.
- Have a clear positive impact on the economy, environment, the society, and more.
- Satisfy one or more of the properties of being flexible in functionalities, scalable in terms of number of inputs and outputs, portable across platforms, efficient in performance, affordable as compared to similar systems, efficient in power consumption, etc.

## 5.2 Model Application Setup in Higher Education

The proposed IDM is adopted by 10 Capstone Design Projects (CDPs) [1-3, 26-32], of which development samples are included in this Chapter. The CDPs are part of an ABET-Accredited Undergraduate Computer Engineering program. The CDP duration spans over one full Academic Year. The system developments start by completing the CRP, DTs, RCL, MAMs, then the AOD and the UCD. The EI Key Indicators (See Table 1) enables the early integration of the evaluation requirements in the design and implementation stages through completing the remaining IDM sub-models. The final product is examined by a team of experts in the field and mainly includes university professors and experts from the industry.

The targeted application setup includes stages that are distributed over two semesters—as follows [20]; the supported steps by the IDM are mapped to the sub-models:

- Semester 1:
  - Problem definition and objectives **[CRP]**

- Project management
- Literature survey **[CRP]**
- Design alternatives and methodology **[DTs]**
- Design specifications **[AOD, UCD]**
- Budgeting **[RCL]**
- Modeling and analysis **[RCL, MAMs, EIG]**
- Prototyping
- Documentation **[All Sub-models]**
- Presentation **[All Sub-models]**

• Semester 2:
  - Implementation **[DTs, AOD, UCD]**
  - Testing and verification
  - Critical appraisal **[RCL, MAMs, EIG]**
  - Documentation **[All Sub-models]**
  - Final product demonstration
  - Presentation **[All Sub-models]**

## *5.3 Model Effectiveness*

The effectiveness of the proposed IoT model is evaluated according to four criteria. Firstly, the opinion of members of the design teams as taken during a focus group. The discussion focused on rating several measures that includes clarity, conciseness, ease-of-use, improving productivity, resolving ambiguity, successful integration of the non-functional requirements and evaluation indicators in the design process, and the model adequacy for academic and industrial applications. The rating is based on five scale points that ranges from *Strongly Disagree* to *Strongly Agree* for the question “To what extent you agree that the IDM achieves the following purpose.” The analysis per measure is presented in Table 2; the presented responses are the averages of scores. In all, participants strongly agree that the proposed development model is clear, concise, easy-to-use, and enables effective IoT system design for students.

The second criteria for investigating the effectiveness of the proposed model is the level of attained scores of the 10 developed systems as examined by a committee of experts. The results show that almost all the developed systems attain scores of 4 and above when mapped onto the EIG. The results are presented in Figure 16.

The 10 CDPs that adopts the IDM demonstrate wide acceptance and appearance in exhibitions and competitions (third criteria); and are published as papers in con-ference proceedings (fourth criteria). The CDPs in [1, 29-31] successfully participated in an Annual Electrical and Computer Engineering Exhibition; [32] received the Student Choice Award, while [30] received the Best Project and Student Choice Awards. The CDPs in [1-3,30,32] successfully led to publishing papers in international conference proceedings.

**Table 2. The average results of the focus group discussions. The scale points are Strongly Disagree (SD), Disagree (D), Neutral (N), Agree (A), Strongly Agree (SA).**

| Measure | SD | D | N | A | SA |
|---|---|---|---|---|---|
| Clear | | | | × | |
| Concise | | | | × | |
| Easy-of-Use | | | | | × |
| Improves productivity | | | | | × |
| Resolves ambiguities | | | | | × |
| Enables the integration of evaluation key indicators | | | × | | |
| Enables effective system design for students | | | | | × |
| Enables effective system design for professionals | | | | × | |

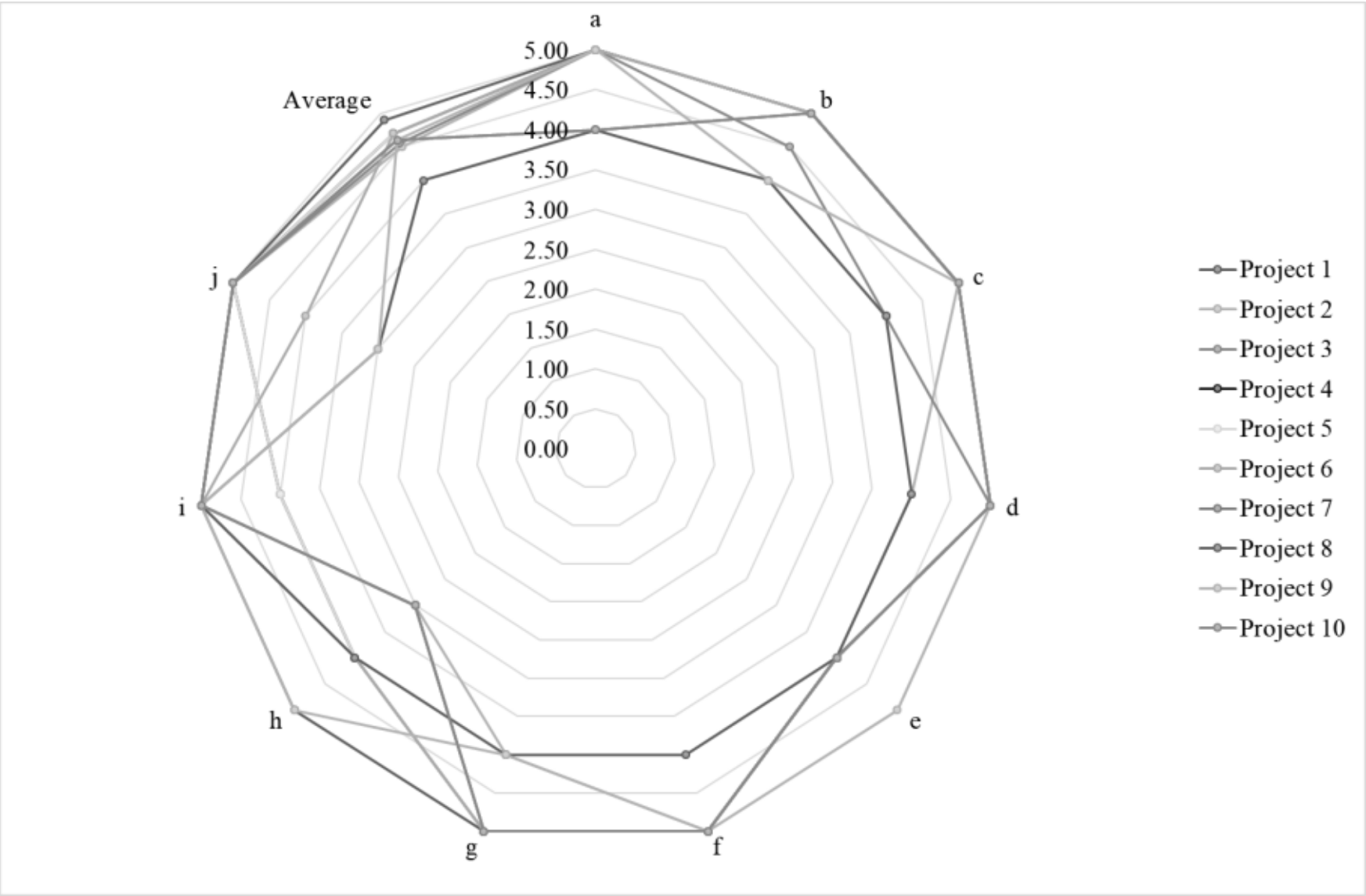

**Fig. 16. The EIG of 10 systems that adopted the IDM**

## *5.4 Closely Related Work*

Several similarities are identified when comparing ARM [6-8] and COMFIT [9] with the IDM. Both approaches and the IDM are used to enable the creation of IoT systems. The three approaches provide high levels of abstraction to the IoT system development. In addition, the three approaches evaluate the model usability and effectiveness in application. On the other hand, ARM clearly targets the business sector as well as the educational sector. Moreover, ARM provides the generation of architectures for specific systems. The benefit of such a generation include providing intrinsic interoperability of the derived IoT systems. COMFIT's high abstraction models are tailored to either the application perspective or the network perspective, thus creating a separation between these two concerns. COMFIT's development environment supports code generation, simulations, and code compilation of applications.

Several differences are noted among the three approaches. Unlike ARM and COMFIT, the IDM lacks security mechanisms. Moreover, the IDM doesn't support automated generation or validation options like COMFIT. COMFIT presents a model validation instance that can help the developers to do such verification during design time and using high-level abstraction models. The verification process, included in COMFIT, can verify inconsistencies and errors in the models. The importance of business goals and the system's requirement are better highlighted in ARM and COMFIT than in the IDM. In Table 3, a comparison among COMFIT, ARM, and IDM in terms of common requirements [8] and supported aspects is presented. COMFIT is found to be superior over the other approaches by its implementation environment, automatic generation capabilities, and validation support. ARM is superior in its support of security mechanisms. Indeed, IDM is superior in integrating mobile solutions; addressing realistic constraints, analysis metrics, and evaluation key indicators; and being mainly aimed for academia. System concept and functionalities are made clearer and more specific by using the CRP, AOD, and UCDs. Unlike ARM and COMFIT, IDM specifies wide technical particularities of an IoT system—from system concept to prototyping options.

**Table 3. Comparison among COMFIT, ARM, and IDM. The symbol "×" means a partial or full support of a requirement or an aspect.**

| Requirement or Aspect | COMFIT | ARM | IDM |
|---|---|---|---|
| Interoperability | × | × | × |
| System development scalability | × | × | × |
| Security mechanisms support | | × | |
| Coverage of different phases of development life cycle | × | × | × |
| Automatic generation support | × | | |
| Design support | × | × | × |
| Implementation environment support | × | | |
| Validation support | × | | |
| Aimed for industry | × | × | |
| Aimed for academia | × | | × |
| Integration of analysis metrics and evaluation indicators | | | × |
| Integration of mobile solutions | | | × |

## *5.5 Limitations and Future Work*

Through the various evaluations of the IDM effectiveness, and comparisons with similar works, a few limitations are noted. The main limitation of the IDM usability is that it is mainly education-oriented rather and limited when it comes to large-scale industrial systems. Moreover, the IDM is informal with no supported validation options during the design process. Furthermore, IDM doesn't address business-related goals in the system requirements. During the focus group discussions, it is noted that the IDM enables limited support to software development; in addition, the evaluation key indicators should be further explained using analytic rubrics.

Work-in-progress and future work include, but are not limited to, expanding the IDM capabilities to target business sector as well as the educational sector. Moreover, improvements include expanding the AOD to include security mechanisms. Future work includes investigating the support of validation during the design time, such as the ability to verify inconsistencies and errors in models and supporting partial automated generations within an IDE.

# 6 Conclusions

The advancement in IoT systems is greatly evident, nowadays, and can be witnessed in almost all aspects of life. The world is becoming closely interconnected through powerful IoT devices that provide true pervasive computing experiences. The aim of this investigation is to present the IDM as an abstract IoT system development model. The IDM comprises several sub-models, such as, the CRP, DTs, RCLs, AODs, CIs, UCDs, MAMs, and EIG. The proposed IDM integrates hardware and software sub-systems, and their interfaces; it enables the design, implementation, analysis, and evaluation, and presentation of an IoT system. Although the model can be adopted by small-scale industrial systems, the IDM in its current form is mainly intended for academic and small-scale use. The IDM is proven to be clear, concise, productive, and effective in application. A variety of applications adopted the IDM, including AgriSys, NFC Wallet, and RECON. In comparisons with similar works, IDM is found to be effective in integrating mobile solutions and addressing realistic constraints, analysis metrics, and evaluation key indicators. Future work includes the creation of an IDE that adopts the IDM, enables automated generations, and supports model validations.